\documentclass[twocolumn,twocolappendix]{aastex701}

\usepackage[varvw,scale=1.08]{newtx}
\usepackage[T1]{fontenc}

\graphicspath{{figures/}}
\usepackage{lipsum}
\usepackage{enumitem}
\usepackage{physics}
\usepackage{multirow}
\usepackage{textgreek}
\usepackage{bm}

\hypersetup{pdfnewwindow=true,
     colorlinks=true,
     linkcolor=blue,
     citecolor=blue,
	 final=true}
	 
\usepackage{etoolbox}
\makeatletter
\patchcmd{\NAT@citex}
  {\@citea\NAT@hyper@{%
     \NAT@nmfmt{\NAT@nm}%
     \hyper@natlinkbreak{\NAT@aysep\NAT@spacechar}{\@citeb\@extra@b@citeb}%
     \NAT@date}}
  {\@citea\NAT@nmfmt{\NAT@nm}%
   \NAT@aysep\NAT@spacechar\NAT@hyper@{\NAT@date}}{}{}
\patchcmd{\NAT@citex}
  {\@citea\NAT@hyper@{%
     \NAT@nmfmt{\NAT@nm}%
     \hyper@natlinkbreak{\NAT@spacechar\NAT@@open\if*#1*\else#1\NAT@spacechar\fi}%
       {\@citeb\@extra@b@citeb}%
     \NAT@date}}
  {\@citea\NAT@nmfmt{\NAT@nm}%
   \NAT@spacechar\NAT@@open\if*#1*\else#1\NAT@spacechar\fi\NAT@hyper@{\NAT@date}}
  {}{}
\makeatother

\newcommand{\mambo}{MAMBO}
\newcommand{\JWST}{\textit{JWST}}
\newcommand{\HST}{\textit{HST}}
\newcommand{\bbarolo}{$^{3{\rm D}}$\textsc{Barolo}}

\newcommand{\hii}{H\,\textsc{ii}}

\newcommand{\cii}{[C\,\textsc{ii}]}
\newcommand{\oiii}{[O\,\textsc{iii}]}

\newcommand{\Lcii}{$L_{[{\rm C}\,\textsc{ii}]}$}
\newcommand{\SFRcii}{${\rm SFR}_{[{\rm C}\,\textsc{ii}]}$}

\newcommand{\um}{\textmu m}

\begin{document}

\title{\large \textit{JWST}+ALMA reveal the ISM kinematics and stellar structure of \\ MAMBO-9, a merging pair of DSFGs in an overdense environment at $\bm{z=5.85}$}

\shortauthors{Akins et al.}
\shorttitle{}

\correspondingauthor{Hollis B. Akins} 
\email{hollis.akins@gmail.com}

\author[0000-0003-3596-8794]{Hollis B. Akins}
\altaffiliation{NSF Graduate Research Fellow}
\affiliation{Department of Astronomy, The University of Texas at Austin, 2515 Speedway Blvd Stop C1400, Austin, TX 78712, USA}
\email{hollis.akins@gmail.com}

\author[0000-0002-0930-6466]{Caitlin M. Casey}
\affiliation{Department of Physics, University of California, Santa Barbara, CA 93106, USA}
\affiliation{Department of Astronomy, The University of Texas at Austin, 2515 Speedway Blvd Stop C1400, Austin, TX 78712, USA}
\affiliation{Cosmic Dawn Center (DAWN), Denmark}
\email{cmcasey@ucsb.edu}

\author[0000-0002-6184-9097]{Jaclyn B. Champagne}
\affiliation{Steward Observatory, University of Arizona, 933 N Cherry Ave, Tucson, AZ 85721, USA}
\email{jbchampagne@arizona.edu}

\author[0000-0003-3881-1397]{Olivia Cooper}
\altaffiliation{NSF Graduate Research Fellow}
\affiliation{Department of Astronomy, The University of Texas at Austin, 2515 Speedway Blvd Stop C1400, Austin, TX 78712, USA} 
\email{ocooper@utexas.edu}

\author[0000-0002-3560-8599]{Maximilien Franco}
\affiliation{Universit\'e Paris-Saclay, Universit\'e Paris Cit\'e, CEA, CNRS, AIM, 91191 Gif-sur-Yvette, France}
\email{francomaximilien.astro@gmail.com}

\author[0000-0002-7530-8857]{Seiji Fujimoto}
\altaffiliation{Hubble Fellow}
\affiliation{David A. Dunlap Department of Astronomy and Astrophysics, University of Toronto, 50 St. George Street, Toronto, Ontario, M5S 3H4, Canada}
\affiliation{Dunlap Institute for Astronomy and Astrophysics, University of Toronto, 50 St. George Street, Toronto, Ontario, M5S 3H4, Canada}
\affiliation{Department of Astronomy, The University of Texas at Austin, 2515 Speedway Blvd Stop C1400, Austin, TX 78712, USA} 
\email{seiji.fujimoto@astro.utoronto.ca}

\author[0000-0002-7821-8873]{Kirsten K. Knudsen}
\affiliation{Department of Space, Earth and Environment, Chalmers University of Technology, SE-412 96 Gothenburg, Sweden}
\email{kirsten.knudsen@chalmers.se}

\author[0000-0002-6610-2048]{Anton M. Koekemoer}
\affiliation{Space Telescope Science Institute, 3700 San Martin Drive, Baltimore, MD 21218, USA}
\email{koekemoer@stsci.edu}

\author[0000-0002-7530-8857]{Arianna S. Long}
\affiliation{Department of Astronomy, The University of Washington, Seattle, WA 98195, USA}
\email{aslong@uw.edu}

\author[0000-0003-2475-124X]{Allison Man}
\affiliation{Department of Physics \& Astronomy, University of British Columbia, 6224 Agricultural Road, Vancouver, BC V6T 1Z1, Canada}
\email{aman@phas.ubc.ca}

\author[0000-0003-0415-0121]{Sinclaire M. Manning}
\altaffiliation{Hubble Fellow}
\affiliation{Department of Astronomy, University of Massachusetts Amherst, 710 N Pleasant Street, Amherst, MA 01003, USA}
\email{smanning@astro.umass.edu}

\author[0000-0002-6149-8178]{Jed McKinney}
\altaffiliation{Hubble Fellow}
\affiliation{Department of Astronomy, The University of Texas at Austin, 2515 Speedway Blvd Stop C1400, Austin, TX 78712, USA}
\email{jed.mckinney@austin.utexas.edu}

\author[0000-0002-7051-1100]{Jorge Zavala} 
\affiliation{Department of Astronomy, University of Massachusetts Amherst, 710 N Pleasant Street, Amherst, MA 01003, USA}
\email{jzavala@umass.edu}

\author[0000-0002-7959-8783]{Pablo Arrabal Haro}
\affiliation{Astrophysics Science Division, NASA Goddard Space Flight Center, 8800 Greenbelt Rd, Greenbelt, MD 20771, USA}
\email{parrabalh@gmail.com}

\author[0000-0001-5414-5131]{Mark Dickinson}
\affiliation{NSF’s National Optical-Infrared Astronomy Research Laboratory, 950 N. Cherry Ave., Tucson, AZ 85719, USA}
\email{mark.dickinson@noirlab.edu}

\author[0000-0002-5588-9156]{Vasily Kokorev}
\affiliation{Department of Astronomy, The University of Texas at Austin, 2515 Speedway Blvd Stop C1400, Austin, TX 78712, USA}
\email{vkokorev@utexas.edu}

\author[0000-0003-1282-7454]{Anthony J. Taylor}
\affiliation{Department of Astronomy, The University of Texas at Austin, 2515 Speedway Blvd Stop C1400, Austin, TX 78712, USA}
\email{anthony.taylor@austin.utexas.edu}

\submitjournal{ApJ}

\begin{abstract} 
	We present high-resolution ALMA \cii\ 158 \um\ observations and \JWST/NIRCam+MIRI imaging of \mambo-9, a pair of optically-dark, dusty star-forming galaxies at $z=5.85$. 
	\mambo-9 is among the most massive, gas-rich, and actively star-forming galaxies at this epoch, when the Universe was less than 1 Gyr old. 
	The new, $400$\,pc-resolution \cii\ observations reveal velocity gradients in both objects; we estimate dynamical masses and find a relative mass ratio of 1:5.
	The kinematics of both objects suggest both rotation and strong tidal interaction, suggesting that the pair has already experienced a close encounter. 
	Indeed, the new \textit{JWST} imaging reveals a continuous bridge of moderately dust-obscured material between the two.
	We perform spatially-resolved SED fitting using the high-resolution ALMA+JWST imaging, finding that the majority of recent star-formation is concentrated in extremely obscured ($A_V > 10$) clouds, while the majority of rest-optical light (stellar continuum and H$\alpha$ emission) is emergent from moderate-to-highly obscured ($A_V\sim 1$--$5$) regions on the outskirts. 
	Combining our new stellar and dynamical mass measurements with previous CO observations, we find that the mass budget of \mambo-9 requires a CO-to-H$_2$ conversion factor ($\alpha_{\rm CO}$) of roughly unity, indicative of a highly metal-enriched ISM. 
	Finally, we show that \mambo-9 resides in a large overdensity spanning the PRIMER-COSMOS field, with 39 galaxies spectroscopically confirmed within $\sim 25$ cMpc. 
	With a total baryonic mass $\sim 10^{11}\,M_\odot$, \mambo-9 can be considered a prototype of massive galaxy formation and likely progenitor of the brightest cluster galaxies (BCGs) in the lower-redshift Universe.
\end{abstract}

\keywords{\small\href{http://astrothesaurus.org/uat/594}{Galaxy evolution (594)}; \href{http://astrothesaurus.org/uat/595}{Galaxy formation (595)}; \href{http://astrothesaurus.org/uat/734}{High-redshift galaxies (734)}; \href{http://astrothesaurus.org/uat/847}{Interstellar medium (847)}; \href{http://astrothesaurus.org/uat/1879}{Circumgalactic medium (1879)}}

\section{Introduction}\label{sec:intro}

The first 3 Gyr of the universe, from the end of reionization around $z\sim 6$ to cosmic noon at $z\sim 2$, marks an era of extraordinary galaxy growth and rapid stellar mass assembly \citep{madauCosmic2014}. 
Much of that growth, however, is obscured by dust; the most intensely star-forming galaxies at $z\gtrsim 2$, the sub-millimeter galaxies (SMGs; \citealt{hughesHighredshift1998, blainSubmillimeter2002}) or dusty star-forming galaxies (DSFGs; \citealt*{caseyDusty2014}), represent the dominant contributor to the obscured star-formation rate density (SFRD).  
To higher redshifts, $z\sim 3$--$6$, these objects can be completely obscured at optical and near-IR wavelengths \citep[e.g.][]{walterIntense2012, riechersDustobscured2013, marroneGalaxy2018, caseyPhysical2019, wangDominant2019, williamsDiscovery2019, taliaIlluminating2021, manningCharacterization2022} and may constitute some $\sim 20$--$50\%$ of the SFRD, which is entirely missed by surveys targeting only the rest-frame UV \citep{zavalaEvolution2021, algeraALMA2023, fujimotoALMA2023}.

The population of DSFGs in the early universe has been elusive. 
Their high obscuration at rest-UV and optical wavelengths has prohibited characterization optical/near-IR surveys, and strong redshift degeneracies with DSFGs at lower-redshift \citep[which are $\sim 20$--$100$ times more numerous, e.g.][]{smolcicMillimeter2012} has made submillimeter selection challenging. 
Much progress has been made on this front in recent years by focusing on gravitationally lensed objects \citep{zavalaConstraining2018, marroneGalaxy2018, sunJADES2023} or observing at longer wavelengths (e.g.~2--3 mm) to efficiently filter-out lower-redshift galaxies \citep[e.g.][]{williamsDiscovery2019, caseyMapping2021, zavalaEvolution2021, cooperSearching2022}. 
	
Moreover, far-infrared emission lines have proven effective tools for redshift confirmation and characterization of high-$z$ DSFGs. 
ALMA spectral line scans have confirmed numerous $z>3$ SMGs via multiple CO and [C\,\textsc{i}] detections \citep{chenALMA2022}. 
At $z\gtrsim 5$, the far-infrared fine-structure lines of \cii\ 158\,\um\ and \oiii\ 88\,\um\ (among others) fall into the submillimeter/millimeter range.
The \cii\ 158\,\um\ line in particular is one of the brightest far-infrared lines, as its low ionization potential ($11.2$ eV) makes it a dominant coolant of the neutral ISM. 
As such, it has served as a powerful tool to measure spectroscopic redshifts \citep[e.g.][]{fudamotoNormal2021, bouwensReionization2022, schouwsALMA2023} gas kinematics 
\citep[e.g.][]{wangStar2013, smitRotation2018, kohandelKinematics2019, jonesALPINEALMA2021, rowlandREBELS252024}, and ISM properties \citep[e.g.][]{harikaneLarge2020, litkeMultiphase2022, litkeISM2023, killiSolar2023}.
Importantly, observations of bright far-IR lines like \cii\ have made possible reliable dynamical mass measurements, even in heavily obscured systems \citep[e.g.][]{rizzoDynamical2021}. 
Regardless, the stellar properties of DSFGs have remained poorly constrained, limiting our ability to place these objects in context in early galaxy evolution.

The launch of the James Webb Space Telescope (\JWST) has opened up our view into the rest-frame optical and near-infrared for $z\gtrsim 4$ galaxies. 
Already, \JWST\ has brought about the first rest-frame optical detections of numerous galaxies previously only detected in the sub-mm \citep[e.g.][]{zavalaDusty2022,mckinneyInfrared2023,sunJADES2023,xiaoAccelerated2024,gentileNotlittle2024,hodgeALESSJWST2025} as well as the identification of previously unknown ``\HST-dark'' massive galaxy candidates at $z\gtrsim 4$ \citep[e.g.][]{barrufetUnveiling2022, barrufetQuiescent2024, chworowskyEvidence2023, williamsGalaxies2023, nelsonFRESCO2023}. 
The high spatial resolution and incredible sensitivity of \JWST\ has also facilitated spatially-resolved analysis, even in heavily dust-obscured objects \citep[e.g.][]{perez-gonzalezCEERS2023, colinaUncovering2023, kokorevJWST2023, crespogomezJWST2024, jonesGANIFS2024, kamieneskiBirds2024, rodighieroOpticallydark2024}. 
These works have generally revealed significant substructure, with strong gradients in stellar and ISM properties across individual galaxies, implying a complex assembly history driven by multiple separate star-formation episodes and/or mergers.

This paper presents high-resolution ALMA and \JWST\ observations of \mambo-9, a pair of massive, dusty star-forming galaxies at $z=5.85$. 
\mambo-9 was first identified by \citet{bertoldiCOSBO2007} with the MAMBO instrument on the Institut de Radioastronomie Millimétrique (IRAM) 30 m telescope, and later corroborated with AzTEC and SCUBA-2 detections \citep{aretxagaAzTEC2011, caseyCharacterization2013}. 
\mambo-9 was identified as a potentially high-$z$ DSFG based on non-detections in \textit{Herschel} imaging at $100$--$500$ \textmu m, and its redshift was confirmed at $z = 5.850$ by \citet{jinDiscovery2019} and \citet{caseyPhysical2019} based on the detection of the $^{12}$CO$(J = 6\to 5)$ and H$_2$O$(2_{1,1}$--$2_{0,2})$ lines.  
Among the highest-redshift DSFGs known, \mambo-9 provides unique insight into the extremes of galaxy formation in the early universe.

Here, we use new high-resolution ($\sim 400$ pc) ALMA \cii\,158\,\textmu m and dust continuum observations, in conjunction with \JWST/NIRCam+MIRI imaging, to constrain \mambo-9's stellar/ISM properties and kinematics. 
In \S\ref{sec:observations} we describe the ALMA and \JWST\ observations and data reduction. 
In \S\ref{sec:results} we describe our main results. 
The the ALMA \cii\ and dust continuum measurements (\S\ref{sec:measurements}) provide kinematics and constraints on the dynamical mass (\S\ref{sec:kinematics}), while the \JWST\ imaging (\S\ref{sec:jwst}), allows derivation of stellar masses from SED fitting (\S\ref{sec:sedfitting}). 
In \S\ref{sec:discussion} we discuss our results, in particular the baryonic mass budget (\S\ref{sec:masses}) and environment of \mambo-9 (\S\ref{sec:overdensity}).  
Throughout this work, we adopt a \citet{kroupaInitial2002} stellar initial mass function and a cosmology consistent with the \citet{planckcollaborationPlanck2020} results ($H_0 = 67.66$\,km\,s$^{-1}$\,Mpc$^{-1}$, $\Omega_{m,0} = 0.31$). 
All magnitudes are quoted in the AB system \citep{okeAbsolute1974}.

\begin{figure*}[t!]
	\centering
	\includegraphics[width=\linewidth]{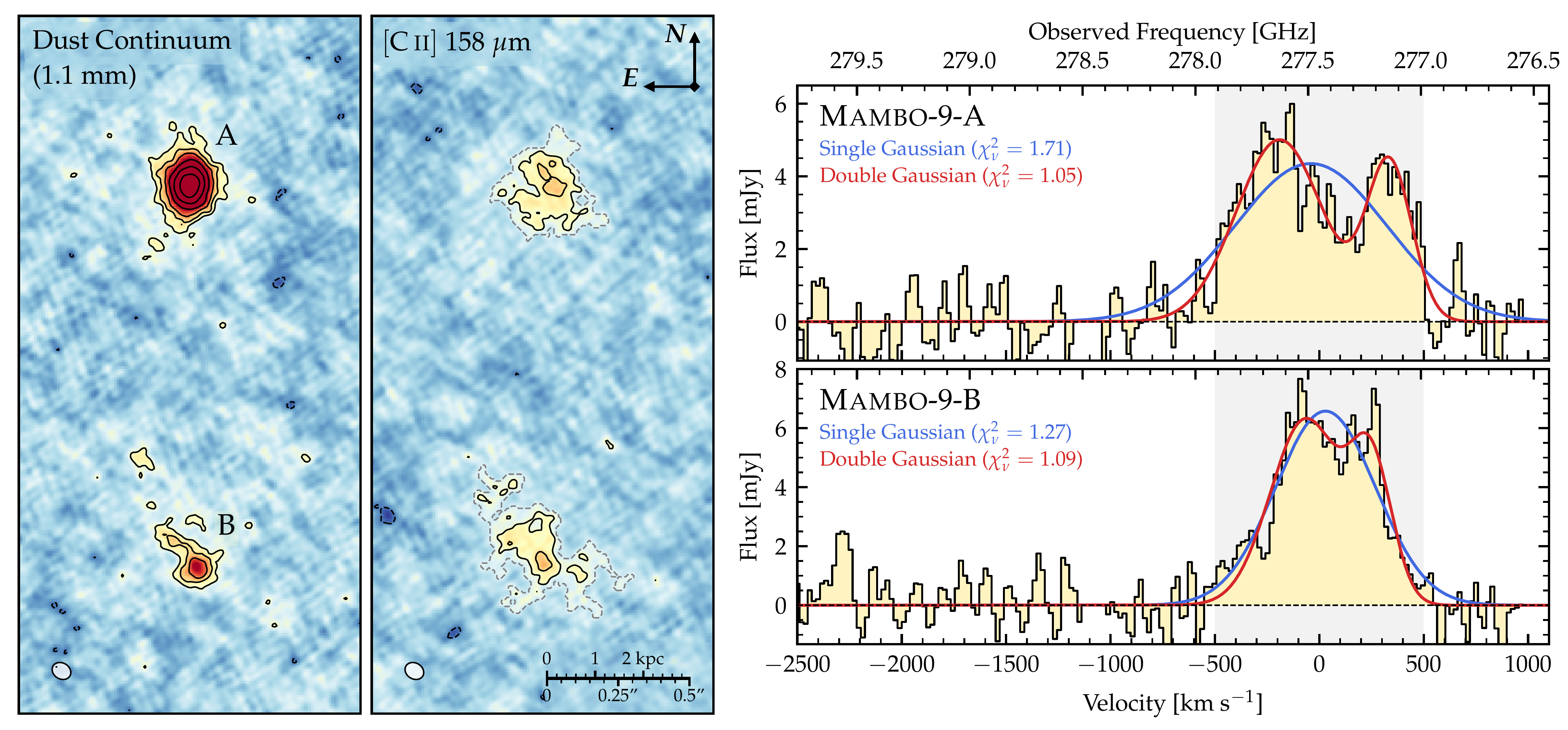}
	\caption{\textit{Left:} ALMA dust continuum and \cii\ 158\,\um\ moment 0 maps for the \mambo-9 system. 
	Contours indicate the significance level, starting from $-3\sigma$, $3\sigma$ and increasing following a Fibonacci sequence thereon. 
	The \cii\ moment 0 map is integrated from $-500 \leq v/{\rm km\,s}^{-1} \leq 500$. 
	The grey dashed contours indicate the segmentation map which is used to extract spectra.
	\textit{Right:} \cii\ line spectra for both \mambo-9-A and B, with gaussian and double gaussian fits applied. 
	For both objects, the double Gaussian fit is a significant improvement over the single Gaussian based on the reduced $\chi^2$ statistic. 
	The double-peaked \cii\ spectra is suggestive of strong rotation, and the dust continuum and \cii\ maps suggest a disturbed morphology due to the galaxy interaction.}
	\label{fig:cont_mom0_spectra}
\end{figure*}

\section{Data and Observations}\label{sec:observations}

\subsection{ALMA Data}\label{sec:data:alma}

We used ALMA observations of \mambo-9 (R.A.~$10^{\rm h}\,00^{\rm m}\,26.358^{\rm s}$, decl. $+02^\circ\,15'\,27.923''$) in band 7 from ALMA program ID \#2021.1.00505.S (P.I.~C.~Casey). 
Based on the spectroscopic redshift of $z=5.85$ \citep{caseyPhysical2019}, the band 7 observations were tuned to cover the \cii~158\,\um\ line; spectral windows were tuned to cover $276.48$--$278.46$, $278.44$--$280.42$, $288.48$--$290.46$, and $290.48$--$292.46$ GHz.
Observations were carried out in the C9 configuration, yielding a nominal beam size of $\sim 0\farcs05$ and a maximum recoverable scale (MRS) of $\sim 0\farcs25''$. 

Reduction and imaging of the ALMA data was performed with the Common Astronomy Software Applications package \citep[CASA;][]{mcmullinCASA2007} version 6.5.2 following the standard ALMA pipeline reduction scripts. 
We additionally identified and flagged a few noisy channels (due to an atmospheric line at $\nu_{\rm obs} \sim 289$ GHz) in one of the continuum spectral windows.
All images were produced with the \texttt{tclean} task in CASA, using natural weighting to maximize sensitivity for any spatially-extended emission given the small MRS of our data.
Cleaning was performed using the ``auto-multithresh'' automasking algorithm \citep{kepleyAutomultithresh2020}, with the parameters recommended in the CASA documentation, and the ``multiscale'' deconvolver \citep{cornwellMultiscale2008} with scales of zero (delta-function) and one times the beam size. 
The continuum map was produced using the multifrequency synthesis \citep[\texttt{mfs};][]{conwayMultifrequency1990} mode in CASA, masking channels $\ge 1000$\,km\,s$^{-1}$ away from the expected line center.
A continuum-subtracted measurement set was produced with the \texttt{uvcontsub} task in CASA, measuring the continuum in the same frequency range with a fit order of 1.
The \cii\ data cube was then produced from this continuum-subtracted measurement set. 

For the continuum map, we obtain an rms of $14$ \textmu Jy beam$^{-1}$. For the \cii\ data cube we obtain an rms of $0.14$ mJy beam$^{-1}$ per $20$ km s$^{-1}$ channel. 
All measurements are done on primary beam corrected maps; we generate an rms map by computing the rms before this correction and weighting by the primary beam.
\mambo-9-A sits at phase center. 
With natural weighting, the synthesized beam size is $0\farcs07 \times 0\farcs06$, which corresponds to a physical resolution (FWHM) of $\sim 400$ pc.

\subsection{JWST Data}\label{sec:data:jwst}

In addition to the high-resolution ALMA data presented in this work, we utilize recent JWST/NIRCam observations taken as part of the Public Release IMaging for Extragalactic Research (PRIMER) survey \citep[GO\#1837, P.I.~J.~Dunlop; see also][]{donnanJWST2024}. 
PRIMER is a large Cycle 1 Treasury Program to image the UDS and COSMOS legacy fields in CANDELS \citep{groginCANDELS2011, koekemoerCANDELS2011} with NIRCam+MIRI. 
PRIMER imaging includes 8 NIRCam bands (F090W, F115W, F150W, F200W, F277W, F356W, F410M, and F444W), plus two MIRI bands (F770W and F1800W).

NIRCam observations covering MAMBO-9 were taken in April 2023, and MIRI observations were taken in November 2023. 
The raw NIRCam imaging was reduced by the JWST Calibration Pipeline version 1.12.1 \citep{bushouseJWST2023},\footnote{\url{https://github.com/spacetelescope/jwst}} with the addition of several custom modifications \citep[as has also been done for other JWST studies, e.g.][]{bagleyCEERS2022} including the subtraction of 1/$f$ noise and sky background.
We use the Calibration Reference Data System (CRDS) pmap 1170 which corresponds to NIRCam instrument mapping imap 0273.
The MIRI/F770W observations were reduced using version 1.12.5 of the JWST Calibration pipeline, along with additional steps for background subtraction that was necessary to mitigate instrumental effects. 
The mosaics in all 10 NIRCam+MIRI bands are produced at a pixel scale of $0\farcs03$/pix and astrometrically aligned to a reference catalog from HST F814W imaging of the field \citep{koekemoerCOSMOS2007}, which is itself aligned to \textit{Gaia} \citep{gaiacollaborationGaia2018}. 
The astrometric accuracy is estimated to be $<5$ (10) mas for NIRCam (MIRI). 
The NIRCam and MIRI imaging reduction follows the methods used for the COSMOS-Web survey \citep[see][]{franco2025, harish2025}.

We additionally make use of recent \JWST/NIRSpec observations from program the CANDELS Area PRISM Epoch of Reionization Survey (CAPERS; GO\#6368, P.I.~M.~Dickinson). 
CAPERS is a \JWST Cycle 3 Treasury program using the NIRSpec MSA with the PRISM to observe very high-redshift galaxies, AGN candidates, and other objects of community interest in three of the CANDELS legacy fields \citep{groginCANDELS2011, koekemoerCANDELS2011}. 
The PRISM disperser covers 0.6--5.3\,$\mu$m at low-resolution ($R\sim 100$). 
CAPERS observed 7 pointings in the PRIMER-COSMOS field, each with 3 separate MSA configurations; high priority targets were observed in in multiple configurations, while lower priority targets were observed in only 1 or 2 to maximize the total spectroscopic yield.
\mambo-9 was selected for follow-up given its high-redshift relative to the full SMG population, and \mambo-9-B received the full 4.74 hours of exposure (\mambo-9-A was not observed). 
Details on the NIRSpec data reduction and optimal extraction is described in \citet{taylorCAPERS2025, donnanCAPERS2025}.

\section{Results}\label{sec:results}

\subsection{High-resolution ALMA \cii\ line and dust continuum measurements}
\label{sec:measurements}

Figure~\ref{fig:cont_mom0_spectra} shows both the high-resolution \cii~158\,\um\ and dust continuum measurements for the \mambo-9 system.
The dust continuum map and \cii~158\,\um\ moment 0 map are shown in the left panels, with contours following a Fibonacci sequence ($-3$, $3$, $5$, $8$, $13$, $21$, $34\sigma$).
The \cii\ moment 0 map is integrated from $-500$ to $+500$ km\,s$^{-1}$ relative to the line center at 277.45\,GHz.  
The right panels show the \cii\ line spectra for the two galaxies with single and double-Gaussian fits applied. 
Spectra and \cii\ flux measurements are extracted from regions within the $>1.5\sigma$ contour on the \cii\ moment 0 map, shown with the grey dashed contours in Figure~\ref{fig:cont_mom0_spectra}.

\mambo-9-A is detected at a peak SNR $\sim 50$ in the dust continuum map and shows a spheroidal/symmetric morphology consistent with previous dust continuum observations \citep{caseyPhysical2019}. 
\mambo-9-B is detected at peak SNR $\sim 13$ in dust continuum in its southwest core, with a prominent tail extending towards the northeast. 
This extended tail is also seen in the \cii\ moment 0 map, though the \cii\ emission is more diffuse. 
The peak position of the \cii\ emission in \mambo-9-A is offset from the peak continuum position, suggesting some decoupling of the \cii\ and IR-emitting gas.

\begin{figure}
	\centering	
	\includegraphics[width=\linewidth]{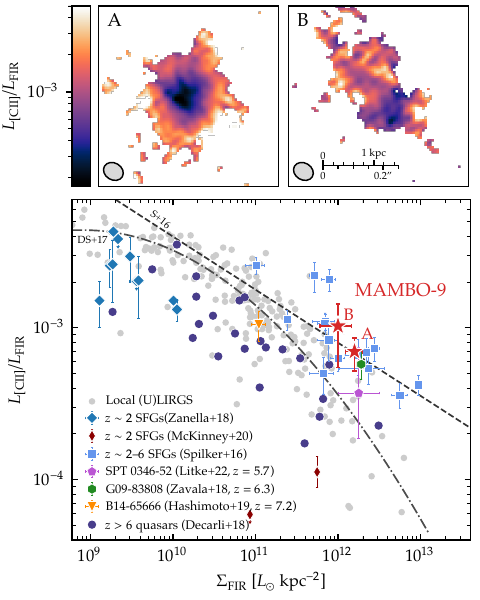}
	\caption{The $L_{[{\rm C}\,\textsc{ii}]}/L_{\rm FIR}$ ratio as a function of $\Sigma_{\rm FIR}$ for the \mambo-9 system and various local and high-redshift galaxies from the literature. We show local ULIRGS from \citep{diaz-santosHerschel2017} $z\sim 2$--$6$ DSFGs from SPT \citep{spilkerALMA2016}, $z\gtrsim 6$ quasars \citep{decarliALMA2018a}, and several individual $z\gtrsim 5$ galaxies \citep{zavalaDusty2018,litkeSpatially2019,hashimotoBig2019}. \mambo-9 follows the trend of decreasing $L_{[{\rm C}\,\textsc{ii}]}/L_{\rm FIR}$ ratio as a function of $\Sigma_{\rm FIR}$ (i.e., the ``\cii\ deficit''). We also show the spatially-resolved $L_{[{\rm C}\,\textsc{ii}]}/L_{\rm FIR}$ ratio maps for \mambo-9-A and B, which follows a similar relationship, indicating that the \cii\ deficit arises from small scales $\lesssim 400$ pc.}
	\label{fig:cii_deficit}
\end{figure}

The \cii\ spectrum for \mambo-9-A shows a double-peaked line profile, separated by $\approx 500$ km\,s$^{-1}$, consistent with the spectrum for $^{12}$CO($J=6\to5$) presented in \citet{caseyPhysical2019}. 
We fit the spectrum to both a single and double-Gaussian model; the latter is a significant improvement over the former, with a reduced chi-squared of $\chi_\nu^2 = 1.05$ vs.~$1.71$. 
The double-peaked emission line spectrum is consistent with strong rotation (see Sec.~\ref{sec:kinematics}), and is also seen in \mambo-9-B, albeit with a smaller separation, suggesting a lower dynamical mass. 
\mambo-9-B also shows a possible third peak at a velocity of $\sim -450$ km s$^{-1}$; however, adding a third Gaussian component does not substantially reduce the $\chi^2_\nu$.

The integrated \cii\ luminosities for \mambo-9-A and B are \Lcii\ $= (3.1 \pm 0.2)\times 10^9~L_\odot$ and $(3.3\pm 0.3)\times 10^9~L_\odot$, respectively.
These luminosities are roughly equal, despite \mambo-9-A being a factor $\sim 3.5$ brighter in the continuum.
Using the \citet{deloozeApplicability2014} calibration for \hii/starburst galaxies, these luminosities would correspond to star-formation rates of \SFRcii\ $=270 \pm 190$ and $290 \pm 200~{\rm M}_\odot$ yr$^{-1}$ for \mambo-9-A and B, respectively. 
These SFRs are underestimated relative to the FIR continuum-derived SFRs \citep{caseyPhysical2019}, which may be due to the ``\cii\ deficit,'' a known phenomenon in which \cii\ emission is reduced in intensely star-forming galaxies. 
Figure~\ref{fig:cii_deficit} shows the \Lcii/$L_{\rm IR}$ ratio as a function of the far-infrared luminosity surface density $\Sigma_{\rm IR}$ for \mambo-9 and various local and high-redshift galaxies. 
In particular, we show local LIRGs \& ULIRGs from GOALS \citep{diaz-santosHerschel2017}, high-$z$ DSFGs from the South Pole Telescope (SPT; \citealt{spilkerALMA2016}), $z\gtrsim 6$ quasars \citep{decarliALMA2018a} and several individual galaxies at $z>5$, including SPT0346-52 at $z=5.66$ \citep{litkeSpatially2019, litkeMultiphase2022}, G09-83808 at $z=6.03$ \citep{zavalaDusty2018}, and B14-65666 at $z=7.2$ \citep{hashimotoBig2019}. 
The IR luminosity is computed from the 1.1\,mm continuum using the best-fit far-IR SED derived in Section~\ref{sec:sedfitting:fir} and integrating over rest-frame $8$--$1000$\,\textmu m.
We also show in the upper panels maps of the \Lcii/$L_{\rm IR}$ ratio for each galaxy, and in the inset panel the distribution of these individual pixels in the \Lcii/$L_{\rm IR}$--$\Sigma_{\rm IR}$ plane. 
\mambo-9-A and B are consistent with the trend of weaker \cii\ in more compact FIR-emitting regions, in both an integrated and spatially-resolved sense. 
Both galaxies show strong \cii\ deficits in their central IR-emitting cores, indicating that the \cii\ deficit arises from small scales $\lesssim 400$ pc, similar to what has been found in other high-$z$ DSFGs \citep{litkeSpatially2019}.

\begin{figure*}
	\centering
	\includegraphics[width=\linewidth]{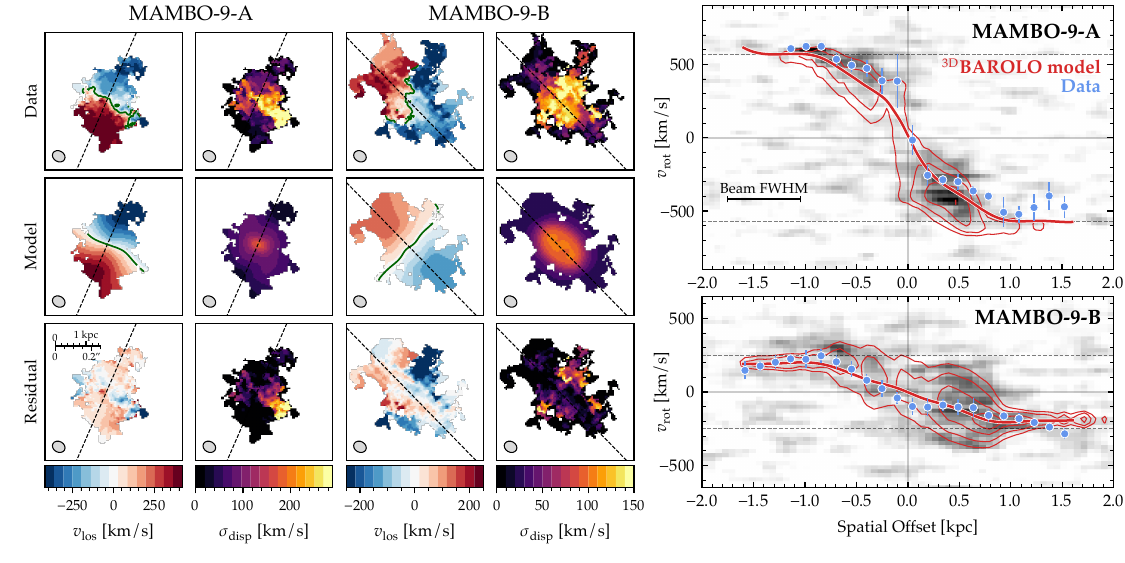}
	\caption{Kinematic modeling of the \mambo-9 system. \textit{Left:} Velocity (moment 1) and velocity dispersion (moment 2) maps for \mambo-9-A and B. We show the data, model, and residual for each moment and object. Dashed lines indicate the kinematic major axis (i.e.~the mean position angle) adopted in \bbarolo\ fits. Green contours show $v=0$ in the moment 1 maps. We adopt $z_A = 5.850$ and $z_B = 5.852$. \mambo-9-A shows clear rotation across its kinematic major axis. Both components show consistently high velocity dispersions ($\sigma \gtrsim 150~{\rm km~s}^{-1}$) even several kpc from the galaxy center.
	\textit{Right:} Major axis position-velocity (PV) diagrams for \mambo-9-A (top) and B (bottom). The background image shows the PV diagrams extracted from cutouts around the two objects, while the red contours show the PV diagram for the best-fit \bbarolo\ model. The rotation curve extracted along the kinematic major axis is shown in blue, and the corresponding model curve is shown in red.}
	\label{fig:mom1_mom2}	
\end{figure*}

Finally, we measure the dust continuum sizes by fitting 2D a S\'ersic profile \citep{sersicInfluence1963} to the ALMA continuum imaging. 
We use the 2D image fitting code \textsc{imfit}\footnote{\url{https://www.mpe.mpg.de/~erwin/code/imfit/}} \citep{erwinIMFIT2015}, in particular the python wrapper \textsc{pyimfit}. 
We convolve the Sérsic model with the PSF, in this case the gaussian clean beam returned by the \textsc{casa} tclean process. 
To provide realistic uncertainties on the model parameters, we utilize Markov Chain Monte Carlo sampling via the \textsc{emcee} python package. 
The 2D Sérsic fits to the ALMA dust continuum images yield Sérsic indicces of $n = 1.71 \pm 0.03$ and $2.9 \pm 0.2$ for \mambo-9-A and B, respectively. 
This suggests that while \mambo-9-A is broadly consistent with a normal disk galaxy, \mambo-9-B has a profile more consistent with a spheroidal galaxy. 
\mambo-9-B is also significantly more extended, with an effective radius of $R_{\rm eff} = 1.17\pm 0.08$ kpc vs.~$469\pm 4$ pc for \mambo-9-A. 
The Sérsic fits yield axis ratios of $b/a = 0.75 \pm 0.01$ and $0.80 \pm 0.02$, which in turn imply disk inclinations of $i = 43.2^\circ \pm 0.5^\circ$ and $38.7^\circ \pm 2.2^\circ$ for A and B, respectively \citep[assuming an intrinsic edge-on axis ratio of 0.25;][]{wuytsKMOS3D2016}.
These statistical uncertainties do not include systematic uncertainties on the relationship between morphological axis ratio and disk inclination; for our dynamical mass measurement, we incorporate systematic effects by combining the morphological and kinematic inclinations (see Section~\ref{sec:kinematics}).
We note also that the Sérsic fit to the dust continuum in \mambo-9-B captures only the bright core in the southwest corner, and not the extended tail to the northeast.

\begin{deluxetable*}{lccCCC}
\centering
\tablecaption{New photometry from \JWST\ and ALMA.\label{tab:photometry}}
\tablehead{Band & \colhead{Wavelength} & \colhead{Units} & \colhead{\mambo-9-A} & \colhead{\mambo-9-B} & \colhead{Bridge}}
\startdata
F090W & 0.9\,$\mu$m & nJy & $<43$ & $<36$ & $<17$ \\ 
F115W & 1.15\,$\mu$m & nJy & $<36$ & $<31$ & $<15$ \\
F150W & 1.5\,$\mu$m & nJy & $<33$ & $<28$ & $<13$ \\
F200W & 2.0\,$\mu$m & nJy & $18\pm 6$ & $<15$ & $6\pm2$ \\
F277W & 2.77\,$\mu$m & nJy & $73\pm3$ & $39\pm3$ & $11\pm1$ \\
F356W & 3.56\,$\mu$m & nJy & $131\pm 3$ & $84\pm3$ & $13\pm1$ \\
F410M & 4.10\,$\mu$m & nJy & $159\pm8$ & $109\pm6$ & $18\pm3$ \\
F444W & 4.44\,$\mu$m & nJy & $227\pm5$ & $189\pm4$ & $19\pm2$ \\
F770W & 7.7\,$\mu$m & nJy & $497\pm14$ & $270\pm11$ & $63\pm6$ \\
F1800W & 18\,$\mu$m & nJy & $886\pm60$ & $1050\pm53$ & $116\pm24$ \\
ALMA-B7 & 1.1\,mm & mJy & $2.7\pm0.1$ & $0.9\pm0.1$ & $<0.1$ 
\enddata
\tablecomments{Non-detections are reported as 3$\sigma$ limits. The ``bridge`` region is indicated in the inset panel of Figure~\ref{fig:SEDfit}.}
\end{deluxetable*}

\subsection{ISM Kinematics}\label{sec:kinematics}
 
The high resolution of the ALMA \cii\ observations facilitate detailed kinematic analysis of both galaxies.
We use the \bbarolo\ software \citep{diteodoro3D2015} to directly fit the datacube to a kinematic model. 
\bbarolo\ models the observed galaxy as a set of concentric, independently rotating rings, approximating a disk. 
The primary advantage of forward-modeling software like \bbarolo, in comparison to direct analysis of position-velocity (PV) diagrams, is that they account for observational effects such as beam-smearing.

The top row of Figure~\ref{fig:mom1_mom2} shows the observed moment 1 and 2 maps (line-of-sight integrated velocity and velocity dispersion) for \mambo-9-A and B. 
These moment maps are extracted from the datacube using the SEARCH algorithm in \bbarolo, iterating to a noise level of 2$\sigma$.
Both galaxies show clear velocity gradients across their major axes (indicated by the dashed black lines). 
At the same time, both galaxies show evidence of disturbed kinematics, with significant asymmetries in the moment 1 map (particularly for \mambo-9-B) and high velocity dispersions ($\sigma \sim 150$--$200$ km/s). 
We note that the systemic velocity for \mambo-9-B is offset slightly from \mambo-9-A, by $\sim 100$\,km\,s$^{-1}$, corresponding to a slightly different redshift of $z=5.852$.

To analyze the PV diagrams and rotation curves, we apply \bbarolo\ tilted-ring modeling to $1.35''$ square cutouts of each galaxy. 
We adopt 10 rings in the models, with a separation of $0\farcs025$, which corresponds to approximately half the resolution of the ALMA observations. 
We assume a thin disk with a scale height of 60 pc.
We conduct a two-stage regularization procedure, first allowing \bbarolo\ to fit all kinematic and geometric parameters (except for the central position, which we determine manually based on the $v=0$ contour line in the moment 1 maps). 
This means that the rotation velocity, velocity dispersion, inclination, and position angle of each ring is fit independently. 
\bbarolo\ then applies parameter regularization and re-fits the data, constraining the inclination and position angle to follow a Bezier function while allowing $v_{\rm rot}$ and $\sigma_{\rm disp}$ to vary freely. 
The derived inclinations range from $44^\circ$--$48^\circ$ for \mambo-9-A and $43^\circ$--$51^\circ$ for \mambo-9-B, similar to, but slightly higher than the inclinations derived from Sérsic fitting.
The middle row of Figure~\ref{fig:mom1_mom2} shows the moment 1 and 2 maps of the best-fit \bbarolo\ model for each galaxy, and the residuals are shown in the bottom row. 
We note that both objects show significant positive residuals in the velocity dispersion, particularly along the kinematic minor axis. 
This may be due to outflowing or kinematically disturbed gas, associated with the close-pass merger.

The rightmost panels of Figure~\ref{fig:mom1_mom2} show the major-axis position-velocity diagrams for both galaxies. 
The ALMA data is shown in grey, and the \bbarolo\ model PV diagram is shown with red contours. 
The blue points show the measured rotation curves extracted from the major axis of the moment 1 maps; the red curve shows the corresponding measurements from the model moment 1 map. 
Rotation velocities are derived from line-of-sight velocities by $v_{\rm rot} = v_{\rm los}/\sin(i)$, where $i$ is the inclination angle. 
Here, we adopt the approximate mean of the morphological and kinematic inclination angles for each galaxy; in particular, we take $i=46\pm 2^\circ$ for \mambo-9-A and $i=45\pm 6^\circ$ for \mambo-9-B.

We derive dynamical masses for each component as
\begin{equation}
  M_{\rm dyn} = \frac{V_{\rm max}^2 R_{\rm max}}{G}
\end{equation}
where $R_{\rm max}$ is the radius of the furthest ring and $V_{\rm max}$ is the maximum rotation velocity. 
We estimate $V_{\rm rot,max} = 570$ km\,s$^{-1}$ for \mambo-9-A and $V_{\rm rot,max} = 250$ km\,s$^{-1}$ for \mambo-9-B. 
Uncertainties in the dynamical mass include the error on the rotation velocity returned by \bbarolo\ as well as the uncertainty on the inclination. 
Since the rotation velocity scales with $1/\sin(i)$ and the dynamical mass scales with $v^2$, an uncertainty on the inclination of $\sigma_i$ would correspond to an additional uncertainty on the dynamical mass of $\sigma_{M_{\rm dyn}}/M_{\rm dyn} = 2\sigma_i \cot(i)$. 
For \mambo-9-A and B, which have fractional uncertainties on $i$ of $\sim 4\%$ and $\sim 13\%$, this contributes an additional $7\%$ and $21\%$ uncertainty to $M_{\rm dyn}$. 
We derive dynamical masses of $(1.1\pm 0.1) \times 10^{11}~M_\odot$ and $(2.0 \pm 0.5) \times 10^{10}~M_\odot$ for components A and B, respectively. 
This corresponds to a merger mass ratio of roughly 1:5, i.e.~a minor merger.

\begin{figure*}
\includegraphics[width=\linewidth]{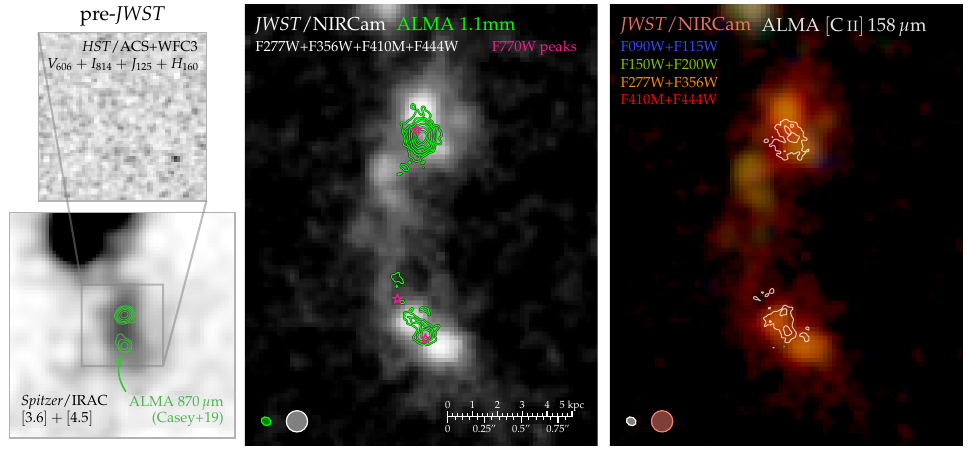}	
\includegraphics[width=\linewidth]{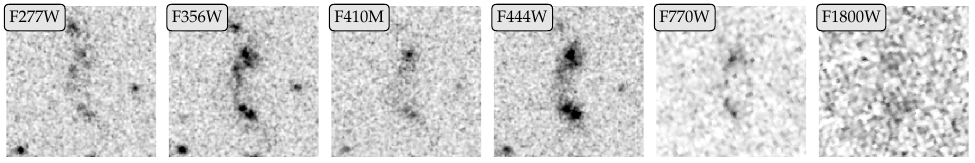}	
\caption{New \JWST\ imaging of \mambo-9, illustrating the utility of \JWST+ALMA for spatially resolving high-$z$ DSFGs. \textit{Top Left:} Stacked cutout of \textit{Spitzer}/IRAC $[3.6]+[4.5]$. \mambo-9 was marginally detected in IRAC imaging, though completely unresolved; identifying the two separate components required targeted ALMA continuum imaging \citep{caseyPhysical2019}. The zoomed inset shows a stacked cutout from combining four HST bands (F606W, F814W, F125W, F160W) from CANDELS \citep{groginCANDELS2011, koekemoerCANDELS2011}, in which \mambo-9 was undetected. \textit{Top Center:} Inverse-variance-weighted stack of the \JWST/NIRCam long wavelength bands, which have been PSF-matched to F444W. Green contours show the high-resolution ALMA imaging of the 1.1mm dust continuum as in Figure~\ref{fig:cont_mom0_spectra}. \textit{Top Right:} False-color RGB image constructed from all eight PSF-matched NIRCam bands. White contours show the \cii~158\,\um\ flux. The resolution of the ALMA data is $\sim 3$ times better than F444W, shown as the ellipses in the bottom-left corner. The bright dust continuum-emitting regions are fainter in the rest-frame optical, suggesting that there is significant obscuration in the core of both galaxies. \textit{Bottom:} \JWST/NIRCam+MIRI imaging of \mambo-9. Panels show 4''$\times$4'' cutouts in F277W, F356W, F410M, F444W, F770W, and F1800W. \mambo-9 is undetected at wavelengths shortward of 2 \um, even in deep \JWST\ imaging.}\label{fig:jwst_imaging}
\end{figure*}

We note that these dynamical mass estimates should still be treated as uncertain, particularly for \mambo-9-B, given the merger state of the two objects and significant turbulent motion present in both.
We measure $V_{\rm rot}/\sigma \approx 4$--$8$ for \mambo-9-A, consistent with a rotation-dominated disk. 
For \mambo-9-B, we measure $V_{\rm rot}/\sigma \approx 0.5$--$3$ in the central part of the galaxy, but $\gtrsim 10$ in the outskirts. 
It is thus unclear if we are seeing a rotating system in \mambo-9-B, or simply tidal effects from the merger. 
We note, however, that it is unlikely that our dynamical mass for \mambo-9-B is \textit{underestimated}---if anything, it is overestimated due to additional motion induced by the merger---supporting that the two galaxies are most likely experiencing a minor merger.

\subsection{JWST/NIRCam+MIRI imaging}
\label{sec:jwst}

Here we present the new \JWST\ imaging of \mambo-9, summarized in Figure~\ref{fig:jwst_imaging}, which provides the first deep, resolved view into the rest-frame optical/near-infrared for this system. 
We first highlight, in the top-left panel, the previous imaging from \HST+\textit{Spitzer}. \mambo-9 is not detected in moderately deep \HST\ imaging from CANDELS \citep{groginCANDELS2011, koekemoerCANDELS2011}, and only marginally detected in \textit{Spitzer}/IRAC imaging at $3.6$ and $4.5$ \textmu m, though unresolved. 
In fact, prior to \JWST, identifying the object as two separate components was only possible with ALMA observations of the dust continuum \citep{caseyPhysical2019}. 
The bottom panels of Figure~\ref{fig:jwst_imaging} show 4''$\times$4'' cutouts around \mambo-9 in 6 \JWST\ bands: F277W, F356W, F410M, F444W, F770W, and F1800W. 
We note that \mambo-9 is undetected in F090W, F115W, F150W, and F200W, even in the moderately deep PRIMER data (F200W $3\sigma$ depth $\sim 28.6$ AB mag), providing upper limits even deeper than the \HST\ non-detections. 
\mambo-9 is faintly detected in F277W (rest-frame $\sim 4000$ \AA), appears clearly in F356W, F410M, F444W, and F770W, and is very marginally detected in F1800W. 
The top-right panels in Figure~\ref{fig:jwst_imaging} show a \JWST/NIRCam LW stack, with the ALMA continuum contours overlaid, and a \JWST/NIRCam RGB image with the \cii\ contours overlaid.

The NIRCam imaging reveals a faint, extended bridge of stellar light between the two components, not visible in the ALMA imaging. 
This bridge is quite red, suggesting it is significantly dust-attenuated or hosts an older stellar population. 
Moreover, it is largely aligned with the kinematic axis of each component (see Section~\ref{sec:kinematics}), suggesting that it may originate from tidal forces induced by the close passage of the two objects.

The MIRI F770W and F1800W bands probe the rest-frame near-IR at $z\sim 6$, which is subject to far less attenuation than the UV/optical. 
Due to the shallow depth of the F1800W imaging, \mambo-9 is only marginally detected, and its morphology is unclear. 
In F770W, however, \mambo-9 is clearly detected and exhibits three clear peaks, one central peak in \mambo-9-A and two in \mambo-9-B. 
The peak F770W positions are slightly offset from the NIRCam stack; to illustrate this, we indicate the positions of the three SNR peaks in F770W on the NIRCam stack in Figure~\ref{fig:jwst_imaging}. 
The F770W peak positions largely align with the ALMA continuum peaks, indicating that the majority of stellar mass arises from the FIR-emitting regions. 
However, at the position of the NE F770W peak in \mambo-9-B, both the ALMA dust continuum and NIRCam imaging are quite faint, indicating perhaps the presence of an older stellar population.

\subsection{JWST/NIRSpec spectrum of \mambo-9-B}\label{sec:nirspec}

\begin{figure}
\includegraphics[width=\linewidth]{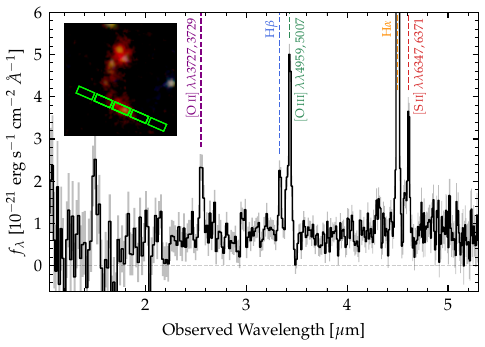}
\caption{\textit{JWST}/NIRSpec PRISM spectrum of \mambo-9-B from CAPERS (GO\#6368). The inset panel shows the NIRCam composite RGB image of \mambo-9 with the NIRSpec slitlet overlaid. We mark the positions of detected emission lines.}	
\end{figure}

As noted in \S\ref{sec:data:jwst}, \mambo-9-B was recently observed with JWST/NIRSpec as part of the CANDELS area PRISM Epoch of Reionization Survey (CAPERS; GO\#6368, P.I.~M.~Dickinson). 
\mambo-9-B was observed with the PRISM disperser (covering 0.6--5.3\,$\mu$m at $R\sim 100$) with an exposure time of 4.74 hrs. 
The spectrum yields detections of $[{\rm O}\,\textsc{ii}]\,\lambda\lambda3727,3729$, H$\beta$, $[{\rm O}\,\textsc{iii}]\,\lambda\lambda4959,5007$, H$\alpha$, and $[{\rm S}\,\textsc{ii}]\,\lambda\lambda6347,6371$. 
We measured line fluxes by fitting Gaussian models to the PRISM spectrum, including a variable local continuum and flexible line width.
We measure a redshift from the PRISM spectrum of $z=5.86\pm 0.01$, consistent with the ALMA \cii\ redshift. 

We measure the nebular dust attenuation from the Balmer decrement. 
We measure line fluxes in H$\alpha$ and H$\beta$ of $5.8^{+0.1}_{-0.1}\times 10^{-19}$ and $7.1^{+1.0}_{-0.9}\times 10^{-20}$ erg\,s$^{-1}$\,cm$^{-2}$\,\AA$^{-1}$. 
The ratio of H$\alpha$/H$\beta$ of $8.3^{+1.2}_{-1.0}$ implies dust attenuation of $A_V = 3.6^{+0.5}_{-0.4}$, assuming a \citet{calzettiDust2001} attenuation law. 
The H$\alpha$ line luminosity (corrected for attenuation) implies a SFR of $19.5^{+7.6}_{-5.4}$\,M$_\odot$\,yr$^{-1}$, using the calibration from \citet{kennicuttStar2012}; this is significantly lower than the SFR derived from \cii\ ($\sim 290$\,M$_\odot$\,yr$^{-1}$) and from the FIR ($\sim 390$\,M$_\odot$\,yr$^{-1}$), perhaps due to the limited physical region probed by the NIRSpec slit or the presence of highly obscured star-forming clouds.

\subsection{SED fitting}\label{sec:sedfitting}

With the new \JWST\ and high-resolution ALMA data, we aim to better physically constrain the stellar properties of the MAMBO-9 system.

\subsubsection{Integrated, Energy-Balance SED fit}\label{sec:sedfitting:energybalance}

To begin, we fit the entire SED from the rest-frame UV to FIR under an energy balance assumption using \textsc{bagpipes} \citep{carnallInferring2018}. 
\textsc{bagpipes} uses the nested sampling algorithm \textsc{Multinest} to conduct Bayesian inference of galaxy physical properties based on user-defined sets of model components. 
We adopt the BPASS library stellar population models \citep{eldridgeBinary2017} and a non-parametric SFH model defined by the $\Delta \log({\rm SFR})$ in adjacent time bins \citep[see][]{lejaHow2019}. 
We construct six fixed age bins at $0$--$10$, $10$--$30$, $30$--$100$, $100$--$200$, $200$--$500$, and $500$--$800$ Myr. 
We use a ``bursty continuity'' prior \citep{tacchellaStellar2022}, e.g. a $t$ distribution with $\sigma=1.0$ and $\nu=2$. 
This prior allows for sharp discontinuities between time bins, in particular, allowing for a recent starburst.
We allow the metallicity to vary from $0$ to $2.5\,Z_\odot$ with a uniform prior. 
Nebular emission is included with $\log U$ allowed to vary from $-4$ to $-1$. 
We adopt a \citet{salimDust2018} parametrization dust attenuation curve with $A_V$ allowed to vary from 0 to 8 and the power-law deviation from a \citet{calzettiDust2000} law $\delta$ from $-0.5$ to $0.5$. 
It include additional dust attenuation around ``birthclouds'' (i.e.,~stars younger than $10$ Myr) via  a multiplicative factor $\eta$, which is allowed to vary from $1$ to $10$ with a uniform prior. 
We found this was necessary to reproduce the extreme SFRs implied by the FIR without significant flux boosting from strong emission lines in the NIRCam bands.
We note that allowing for more dust attenuation around birthclouds means that the mass-weighted $A_V$ can be higher than $8$, the nominal maximum of the prior. 
Finally, dust emission is modeled using the \citet{draineInfrared2007} template set with $q_{\rm PAH} \sim 0.1$--$4.58$, $U_{\rm min} \sim 0.1$--$25$, and $\gamma\sim 0.001$--$0.7$. 
We assume energy balance, such that the total energy absorbed by dust is reradiated in the FIR/sub-mm.

\begin{figure*}
\centering
\includegraphics{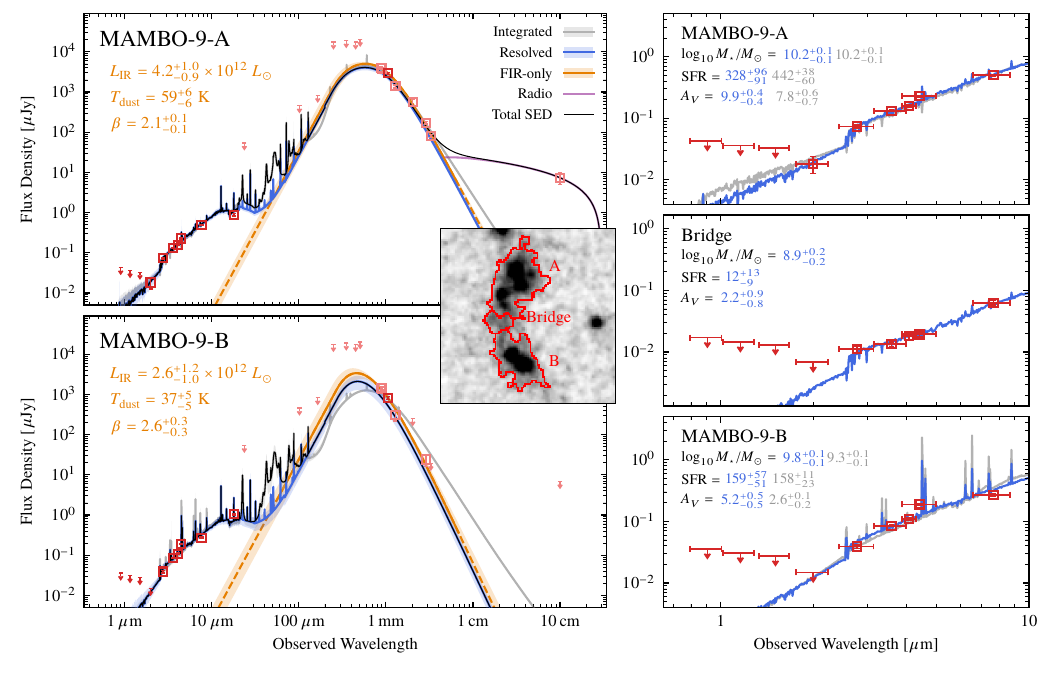}
\caption{Spectral energy distributions and best-fit models for \mambo-9. The FIR-only SED fits are shown in orange, the spatially-resolved energy balance fits are shown in blue, and the integrated fits is shown in grey. Existing photometry as compiled by \citet{caseyPhysical2019} is shown in light red, while new \JWST\ and ALMA photometry from this work is shown in dark red. The right panels show zoom-ins of the optical-near-IR SEDs for \mambo-9-A and B, and the bridge between the two, as extracted from the regions shown in the inset cutout.}\label{fig:SEDfit}
\end{figure*}

Figure~\ref{fig:SEDfit} shows the full panchromatic SED for \mambo-9-A and B; we first highlight the best-fit \textsc{bagpipes} SED fits to the integrated photometry in grey. 
The right panels show zoom-ins to the optical/near-IR portion of the SED. 
The best-fit SED for \mambo-9-A includes a moderate Balmer break, indicating an older stellar population, required by the deep NIRCam/F200W upper limit. 
The SED fit yields a stellar mass of $\log M_\star/M_\odot = 10.2^{+0.1}_{-0.1}$, a star-formation rate ${\rm SFR} = 440^{+40}_{-60}$\,M$_\odot$\,yr$^{-1}$, and a dust-attenuation $A_V = 7.8^{+0.6}_{-0.7}$. 
For \mambo-9-B, the SED fit requires stronger emission lines to reproduce the F444W excess, and yields a stellar mass of $\log M_\star/M_\odot = 9.3^{+0.1}_{-0.1}$, a star-formation rate ${\rm SFR} = 160^{+10}_{-20}$\,M$_\odot$\,yr$^{-1}$, and a dust-attenuation $A_V = 2.6^{+0.1}_{-0.2}$. 
We note that these SFRs are somewhat underestimated relative to those derived from our dedicated FIR-only SED fits (\S\ref{sec:sedfitting:fir}), and particularly for \mambo-9-B, the \citet{draineInfrared2007} dust models do a poor job of fitting the FIR SED. 
As such, we perform dedicated fits to the FIR SED in the following section.

\subsubsection{FIR SED fitting}\label{sec:sedfitting:fir}

We additionally perform dedicated fits to the FIR SED, absent any assumptions about energy balance. 
The existing literature photometry for \mambo-9, from the rest-frame UV to FIR, is summarized in Table 2 of \citet{caseyPhysical2019}; we adopt these MIR and FIR measurements in our SED fitting.
We include the new ALMA band 7 data ($\lambda \sim 1.05$ mm), with measured fluxes of $2.93\pm 0.08$ and $0.82 \pm 0.06$ mJy for \mambo-9-A and B, respectively. 
Note that we omit the unresolved \mambo, AzTEC, and SCUBA-2 850\,\um\ data as the higher-resolution ALMA observations sufficiently cover this wavelength range, and at much higher SNR. 
We include the non-detections from \textit{Spitzer}/MIPS, \textit{Herschel}/PACS+SPIRE, and SCUBA-2 450\,\um\ as upper limits.

We fit the FIR SED to a modified blackbody$+$power law as described in \citet{caseyFarinfrared2012} and \citet{drewNo2022}; that is, the FIR SED follows a modified blackbody in the Rayleigh-Jeans tail and switches to a power law in the Wien tail in order to account for the presence of multiple increasingly hot, increasingly faint dust components. 
In the absence of strong constraints at rest-frame $10$--$100$\,$\mu$m we fix the power law slope $\alpha_{\rm MIR} = 4.0$. 
We adopt an optically-thin blackbody for \mambo-9-B, while for \mambo-9-A we adopt the more general opacity model described in \citet{caseyPhysical2019} in which $\tau=1$ at $\lambda = 200$~\,\um. 
We perform fitting using a Markov Chain Monte Carlo routine with flat priors on $L_{\rm IR}$, $T_{\rm dust}$, and $\beta$, and correct for the effects of heating by and decreasing contrast against the CMB following \citet{dacunhaEffect2013a}.

\begin{figure*}
	\centering
	\includegraphics[width=\linewidth]{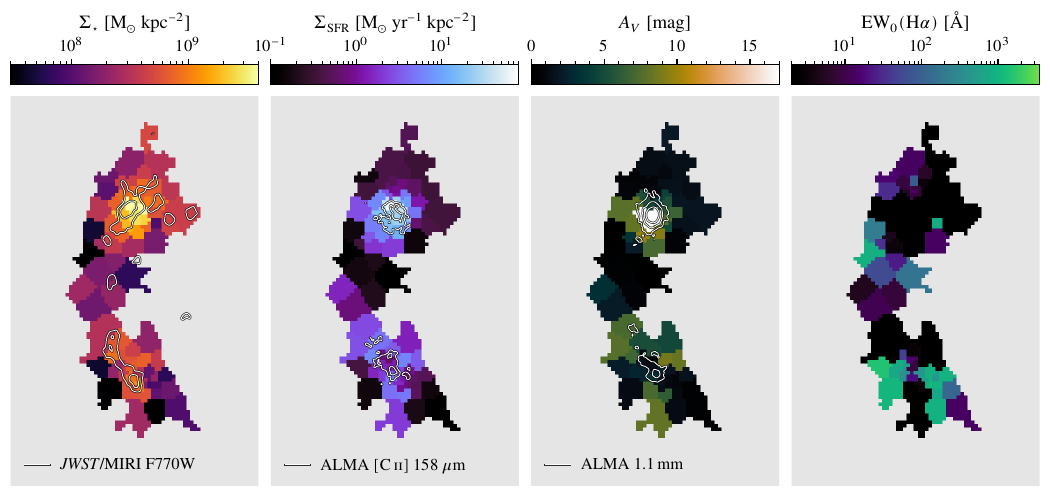}	
	\caption{Results from our spatially-resolved SED fitting procedure. From left to right, we show maps of the best-fit stellar mass surface density $\Sigma_\star$, SFR surface density $\Sigma_{\rm SFR}$ (over the last 100 Myr), dust attenuation $A_V$ (including additional attenuation towards young stars), and equivalent width EW$_0({\rm H}\alpha)$ in each Vonoroi bin.} \label{fig:maps}
\end{figure*}

The resulting FIR-only fits are shown in orange in Figure~\ref{fig:SEDfit}. 
For \mambo-9-A, we derive a dust temperature of $59.0^{+6.3}_{-6.6}$ K and an IR luminosity of $(4.2^{+1.0}_{-0.9})\times 10^{12}~L_\odot$. 
For \mambo-9-B, we derive a dust temperature of $37.9^{+5.7}_{-5.8}$ K and an IR luminosity of $(2.6^{+1.2}_{-1.0})\times 10^{12}~L_\odot$. 
Using the calibration from \citet{kennicuttStar2012}, these IR luminosities correspond to obscured star-formation rates of ${\rm SFR}_{\rm IR} = 620^{+150}_{-130}$ and $390^{+180}_{-150}~M_\odot~{\rm yr}^{-1}$ for \mambo-9-A and B, respectively. 
These are slightly different from, but consistent with the measurements from \citet{caseyPhysical2019}, and notably higher than the SFRs inferred from our integrated SED fits.

\subsubsection{Spatially-resolved SED fitting}

Given the complex multiwavelength morphology presented in Section~\ref{sec:jwst}, with optical/near-IR emission somewhat decoupled from the FIR emission, we additionally perform spatially-resolved SED fitting to capture spatial variations in the dust attenuation and stellar population properties across the two galaxies. 
While the signal-to-noise is not sufficient to perform pixel-by-pixel SED fitting, we construct 2D spatial bins using the \texttt{vorbin} package in Python. 
\texttt{vorbin} takes a 2D image and noise map as input and constructs spatial Voronoi bins to yield a specified target SNR per bin. 
We define the bins based on a composite of the NIRCam LW stack (shown in Figure~\ref{fig:jwst_imaging}) and the ALMA dust continuum map, which has been $uv$-tapered by 0\farcs13 and resampled to match the \JWST\ resolution. 
We set the target SNR to 30, which we find produces a reasonable number of bins (53) with sufficient signal-to-noise in each band to derive physical properties. 
We note that we combined multiple Voronoi bins to construct the extraction apertures for \mambo-9-A and B used in the integrated SED fits in the previous section (shown in the inset panel of Figure~\ref{fig:SEDfit}). 
We refer to the remaining apertures as probing the ``bridge'' between the two objects.

For each bin, we fit the SED to the \JWST/NIRCam+MIRI photometry and the $uv$-tapered ALMA 1.1\,mm dust continuum map; we use the same stellar, nebular, and dust attenuation priors from our integrated \texttt{bagpipes} run. 
However, we model the dust emission using a modified blackbody+power law as in Section~\ref{sec:sedfitting:fir}, rather than the \citet{draineInfrared2007} templates, as this provides a better fit to the data, particularly for \mambo-9-B. 
We require energy balance within each bin, and given that we have only one spatially-resolved data point in the FIR, we reduce the number of free parameters in the model by fixing $T_{\rm dust}$ and $\beta$ to their best-fit values from the FIR-only SED fits.

Figure~\ref{fig:maps} shows the results of our spatially-resolved SED fitting procedure. 
In particular, we show the stellar mass surface density $\Sigma_\star$, SFR surface density $\Sigma_{\rm SFR}$, dust attenuation $A_V$, and inferred H$\alpha$ equivalent width in each 2D bin. 
For \mambo-9-A, the bulk of the stellar mass and recent star-formation is concentrated in the central core, which is highly obscured ($A_V > 10$). 
\mambo-9-B, by contrast, is more moderately attenuated ($A_V\sim5$) and exhibits high H$\alpha$ equivalent widths (EW$_0 \gtrsim 100$\,\AA).
These values are slightly in excess of those inferred form the NIRSpec spectrum presented in \ref{sec:nirspec}, which shows an H$\alpha$ equivalent width of EW$_0 = 93^{+6}_{-6}$~\AA\ and $A_V = 3.6^{+0.5}_{-0.4}$; however, this could be due to the different physical regions probed by the NIRSpec slit and the resolved SED fit. 
In the future, deep IFU spectroscopy could be used to more robustly examine the spatial variation in the dust attenuation, as well as other properties, such as gas-phase metallicity.

The final SEDs for \mambo-9-A and B, constructed by summing the SEDs for each spatial bin, are shown in blue in Figure~\ref{fig:SEDfit}. 
For \mambo-9-A, we derive a stellar mass of $\log M_\star/M_\odot = 10.2^{+0.1}_{-0.1}$, a star-formation rate ${\rm SFR} = 330^{+100}_{-100}$, and a (mass-weighted) dust-attenuation $A_V = 9.9^{+0.4}_{-0.4}$; these values are consistent with those inferred from the simpler integrated SED fit. 
For \mambo-9-B; we derive  $\log M_\star/M_\odot = 9.8^{+0.1}_{-0.1}$, ${\rm SFR} = 160^{+60}_{-50}$, and $A_V = 5.2^{+0.5}_{-0.5}$; the stellar mass and dust attenuation are somewhat large than in the integrated SED fit, likely due to the better accounting of energy-balance using the MBB+PL dust SED. 
We consider these results as our fiducial measurements of the stellar masses in both components.

\section{Discussion}\label{sec:discussion}

\subsection{A highly dust-obscured merger at $z=5.85$}\label{sec:nature} 

\mambo-9 is among the highest-redshift spectroscopically-confirmed DSFGs, and provides valuable insight into the extreme star-formation conditions of massive galaxies in the first Gyr. 
From our spatially-resolved SED fitting, we derive a high stellar mass density in the central region of \mambo-9-A, $\log \Sigma_{\star}/(M_\odot\,{\rm kpc}^{-2}) \gtrsim 9.5$, similar to DSFGs and massive quiescent galaxies at cosmic noon \citep[e.g.][]{suessDissecting2021}. 
Moreover, we find that the bulk of the recent star-formation, traced by the dust and \cii\ emission, originates from the compact central core of each object, from clouds with very high dust attenuation $A_V \gtrsim 10$. 
This is similar to what is seen in the nuclear region of Arp 220, a local ULIRG comprised of two merging starbursts each with highly concentrated nuclear disks \citep{scovilleALMA2017}.
Estimates of the nuclear attenuation in Arp 220 range from $A_V \sim 6$ to $\sim 100$ \citep{haasPAH2001, engelArp2011}, with recent measurements from \JWST/NIRSpec $\sim 11$--$14$ \citep{pernaNo2024}.   
Recent spatially-resolved studies using \JWST\ have found potentially substantial optically-thick substructures in relatively normal, massive galaxies at $z\sim 3$, suggesting that this could be a common feature in massive galaxies \citep{Cheng2025}. 
Future \JWST\ spectroscopy could confirm the high dust attenuation in \mambo-9, via direct measurements of the Balmer and/or Paschen line decrements.

The \JWST\ imaging has revealed a continuous bridge of material connecting \mambo-9-A and B, likely a tidal tail from a recent close pass between the two objects. 
This would be among the highest-redshift tidal tails detected \citep[though, see][]{sugaharaRIOJA2024}.
Interestingly, the tidal tail is itself quite red, and moderately dust-obscured ($A_V\sim 1$), indicating significant dust enrichment even $\sim 4$ kpc from the central SF core(s). 
Moreover, the tidal tail/bridge between the two objects comprises a few $\times 10^9~M_\odot$ of stellar mass, $\sim 7\%$ of the total stellar mass of the \mambo-9 system.

We note that the new MIRI F1800W data do not provide any evidence for a buried AGN in \mambo-9, which would provide a strong MIR flux enhancement \citep[e.g.][]{ivisonSpitzer2004}.
Though both components are only marginally detected, the measurements are 2 orders of magnitude below \textit{Spitzer}/MIPS 24\,$\mu$m limits. 
The SEDs of \mambo-9-A and B are consistent with stellar emission dominating at rest-frame $\sim 2$--$3$\,$\mu$m, with no need for a significant AGN torus component.

\begin{figure}
\centering
\includegraphics[width=\linewidth]{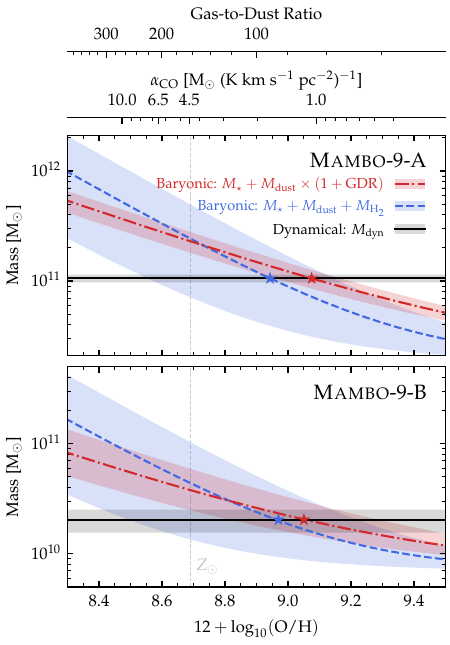}	
\caption{Mass budget of \mambo-9-A and B as a function of the metallicity used to compute the gas mass. Red/blue lines and shaded regions indicate the total baryonic mass ($M_\star + M_{\rm dust} + M_{\rm gas}$), computed assuming either an $\alpha_{\rm CO}$ (blue, dashed line) or a gas-to-dust ratio (red, dot-dash line), adopting standard calibrations with metallicity for both (see text). Note that the CO-based gas mass includes the uncertainty on the CO SLED. The black line and shaded region indicates the dynamical mass constraint derived in this work. The new constraints on the stellar and dynamical masses require $\alpha_{\rm CO}\sim 1$--$2$ or ${\rm GDR}\sim 70$--$90$, consistent with super-solar metallicity.}\label{fig:mass_budget}
\end{figure}

\subsection{Mass Budget}\label{sec:masses} 

The \cii\ kinematics and broadband SED fitting presented in this work provide independent constraints on the dynamical and stellar masses, respectively, of the \mambo-9 system. 
Here we incorporate existing constraints on the gas and dust mass to discuss the mass budget of \mambo-9, with particular focus on the conversion factors to derive gas masses. %

First, we consider the $L_{\rm CO}$ to $M_{{\rm H}_2}$ conversion factor, $\alpha_{\rm CO}$. 
While the MW ratio of $\alpha_{\rm CO} \sim 4.5$ M$_\odot$~$({\rm K~km~s}^{-1}\,{\rm pc}^{-2})^{-1}$ (6.5 including He gas) is commonly adopted, the conversion factor is known to scale with metallicity. 
Here we assume the Helium-corrected $\alpha_{\rm CO}$ scaling from \citet[][their Eq.~25]{accursoDeriving2017}, ignoring the dependence on SFR, which is marginal compared to the dependence on metallicity. 
We adopt $L'_{\rm CO(6-5)} = (1.4\pm 0.9)\times 10^{10}$ K km s$^{-1}$ pc$^{-2}$ for \mambo-9-A and $(2.3\pm 0.7)\times 10^{9}$ K km s$^{-1}$ pc$^{-2}$ for \mambo-9-B \citep{caseyPhysical2019}.
We assume that the CO spectral line energy distribution (SLED) follows the average for high-$z$ DSFGs, such that $I_{{\rm CO}(6-5)}/I_{{\rm CO}(1-0)} = 10^{+30}_{-5}$ \citep*{caseyDusty2014}, i.e.~$L_{{\rm CO}(1-0)}' \sim 3.6L_{{\rm CO}(6-5)}'$. 
We then calculate the H$_2$ mass as $\alpha_{\rm CO} L_{{\rm CO}(1-0)}'$. 

Second, provided there are observations on the Rayleigh-Jeans tail, gas masses can also be estimated from dust masses assuming a gas-to-dust ratio (GDR) \citep[e.g.][]{scovilleISM2016}, which is known to also scale with metallicity. 
We adopt the dust masses measured in \citet{caseyPhysical2019}: $(1.3\pm 0.3)\times 10^9~M_\odot$ and $(1.9^{+1.3}_{-0.8})\times 10^8~M_\odot$ for components A and B, respectively, which are measured from 3 mm continuum imaging. 
We assume the GDR scaling with metallicity from \citet{remy-ruyerGasdust2014}.

Figure~\ref{fig:mass_budget} shows the mass budget of \mambo-9-A and B as a function of the metallicity assumed in computing $\alpha_{\rm CO}$ and the GDR, which are shown in the two additional axes at the top. 
The dynamical mass is shown as a black horizontal line, and the total baryonic mass ($M_\star + M_{\rm dust} + M_{\rm gas}$) is shown in blue and red for gas mass calculations based on $\alpha_{\rm CO}$ and GDR, respectively. 
We mark with stars the point at which the total baryonic mass matches the dynamical mass.

The dynamical mass measurements require gas masses of $9.1^{+3.6}_{-2.7}\times 10^{10}$\,M$_\odot$ for \mambo-9-A and $1.3^{+1.0}_{-0.7}\times 10^{10}$\,M$_\odot$ for \mambo-9-B (see Table~\ref{tab:properties}). 
This corresponds to $\alpha_{\rm CO} \sim 1$--$2$, consistent with other dusty star-forming galaxies \citep{downesRotating1998, tacconiSubmillimeter2008, genzelStudy2010, bothwellSurvey2013, zavalaProbing2022} and in stark contrast to expectations for low-metallicity environments, where $\alpha_{\rm CO}$ can reach $\sim 100$ \citep{maddenTracing2020, boogaardMeasuring2021a, jiaoCarbonH22021, sandersCO2022}. 
Similarly, our results require a gas-to-dust ratio $\sim 70$--$90$. 
These results are consistent with slightly super-solar metallicity, implying an evolved stellar population/ISM in already by $z=5.85$.

\begin{deluxetable}{lccc}
\tablecolumns{4}
\tablecaption{Measured and derived properties for \mambo-9.\label{tab:properties}}
\tablehead{Property & Units & \mambo-9-A & \mambo-9-B}
\startdata
$z$ & \dots & 5.8500 & 5.8518 \\
R.A. & hms & 10:00:26.358 & 10:00:26.357 \\
Decl. & dms & +02:15:27.92 & +02:15:26.59 \\[-0.5em] 
\sidehead{\it From ALMA FIR continuum measurements\dotfill}\\[-1.8em]
$T_{\rm dust}$ & K & $59^{+6}_{-6}$ & $37^{+5}_{-5}$ \\
$\beta$ & \dots & $2.1^{+0.1}_{-0.1}$ & $2.6^{+0.3}_{-0.3}$ \\
$L_{\rm IR}$ & $10^{12}~L_\odot$ & $4.2^{+1.0}_{-0.9}$ & $2.6^{+1.2}_{-1.0}$ \\
$\Sigma_{\rm IR}$ & $10^{11}~L_\odot~{\rm kpc}^{-2}$ & $15.8^{+3.9}_{-3.4}$ & $10.0^{+4.6}_{-3.9}$ \\
${\rm SFR}_{\rm IR}$ & $M_\odot~{\rm yr}^{-1}$ & $620^{+150}_{-130}$ & $390^{+180}_{-150}$ \\
$M_{\rm dust}$$^\dagger$ & $10^8~M_\odot$ & $13\pm 3$ & $1.9^{+1.3}_{-0.8}$ \\
Sérsic $n$ & \dots & $1.71\pm 0.03$ & $2.9\pm 0.2$ \\
$R_{\rm eff}$ & pc & $469\pm 4$ & $1170\pm 80$ \\
$b/a$  & \dots & $0.75\pm0.01$ & $0.80\pm0.02$ \\
$i$ & $^\circ$ & $43.5\pm0.5$ & $38.4\pm2.4$ \\[-0.5em] 
\sidehead{\it From ALMA $[{\rm C}\,\textsc{ii}]~158~\mu$m line measurements\dotfill}\\[-1.8em]
\Lcii  & $10^9~L_\odot$ & $3.1 \pm 0.2$ & $3.3\pm 0.3$ \\
\SFRcii & $M_\odot~{\rm yr}^{-1}$ & $270\pm 190$ & $290\pm 200$ \\
$M_{\rm dyn}$ & $10^{10}~M_\odot$ & $10.6 \pm 0.9$ & $2.0\pm0.5$ \\[-0.5em] 
\sidehead{\it From broadband SED fitting\dotfill}\\[-1.8em]
$M_\star$ & $10^{9}~M_\odot$ & $17.5^{+5.5}_{-4.2}$ & $6.9^{+2.4}_{-1.8}$ \\ 
$A_V$ & mag & $9.9^{+0.4}_{-0.8}$ & $5.2^{+0.5}_{-0.5}$ \\
${\rm SFR}$ & $M_\odot~{\rm yr}^{-1}$ & $330^{+100}_{-100}$ & $160^{+70}_{-60}$ \\[-0.5em]
\sidehead{\it From mass budget analysis\dotfill}\\[-1.8em]
$M_{\rm gas}$ & $10^{10}~M_\odot$ & $9.1^{+3.6}_{-2.7}$ & $1.3^{+1.0}_{-0.7}$ \\
$f_{\rm gas}$ & \dots & $83^{+2}_{-2}\%$ & $64^{+8}_{-12}\%$ 
\enddata
\tablenotetext{\dagger}{Dust masses are taken from \citet{caseyPhysical2019}, derived from 3 mm continuum observations.}
\end{deluxetable}

\subsection{The environment around MAMBO-9}\label{sec:overdensity}

\begin{figure}
	\centering
	\includegraphics[width=\linewidth]{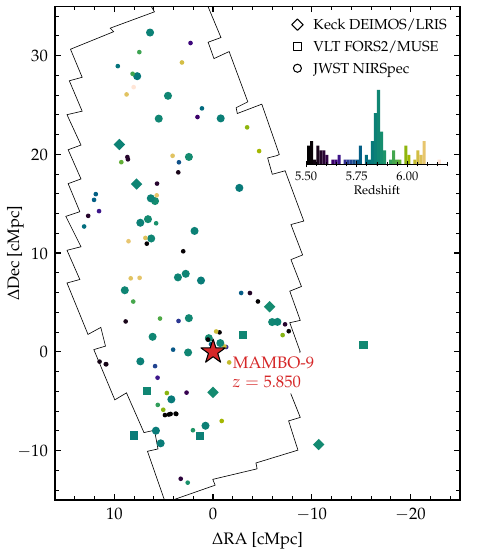}
	\caption{The overdense environment around MAMBO-9.
	We show objects spectroscopically-confirmed in CAPERS between $z=5.5$ and $z=6.2$, objects within $\Delta z = 0.03$ of MAMBO-9 are shown as larger points. 
	We supplement this with 10 objects spectroscopically-confirmed from ground-based Lyman-$\alpha$ surveys on Keck and VLT \citep{hasingerDEIMOS2018,pentericciCANDELSz72018,schmidtRecovery2021,epinatMAGIC2024} as compiled by ; \citet{Khostovan2025}. 
	The inset panel shows the redshift histogram, highlighting the clear excess of sources between $z=5.8$--$5.9$. 
	Note the large spatial scale of the structure, $\sim 40$ cMpc, spanning most of the PRIMER-COSMOS field.}\label{fig:overdensity}	
\end{figure}

Finally, we discuss whether \mambo-9 resides in an overdense environment. 
For many years, it has been suggested that unusually massive and rare galaxies---including quasars \citep{venemansMost2002, venemansProperties2005, venemansProtoclusters2007, shimasakuSubaru2003, kuiperSINFONI2011} and DSFGs \citep{chapmanInterferometric2008, dannerbauerExcess2014, caseyUbiquity2016a, hungLargescale2016, oteoExtreme2018, millerMassive2018, marroneGalaxy2018}---can serve as beacons of the most massive structures still in the protocluster phase before virialization.  
Indeed, cosmological simulations predict massive DSFGs to sit in the progenitor structures of eventual galaxy clusters \citep{millerBias2015, chiangGalaxy2017}. 
Recent \JWST\ results from the JADES \citep{eisensteinOverview2023} and FRESCO \citep{oeschJWST2023} surveys reveal a large-scale overdensity at $z\sim 5.2$ in GOODS-N \citep{sunJADES2023, herard-demancheMapping2023}, around the sub-millimeter galaxies HDF850.1 ($z=5.184$) and GN10 ($z=5.303$). 
Where previous spectroscopic efforts had identified one quasar and $\sim 11$ galaxies at similar redshifts \citep{walterIntense2012}, these recent NIRCam WFSS grism observations have detected $\sim 100$ H$\alpha$ emitters at the same redshift within a (15 cMpc)$^3$ region.

The story of \mambo-9 appears to follow a similar pattern: several Ly$\alpha$ emitters have been serendipitously identified at $z\approx 5.84$--$5.88$ nearby, largely from Keck or VLT spectroscopy \citep{hasingerDEIMOS2018, pentericciCANDELSz72018, schmidtRecovery2021, epinatMAGIC2024}, and another unique optically-dark DSFG, COS-3mm-1 \citep{williamsDiscovery2019} has a tentative CO(6--5) line detection at $z = 5.85$ \citep{zavalaTentative2021} though it is $\sim 0.5^\circ$ ($\sim 70$ cMpc) away from \mambo-9. 
This tentative structure was thought to be related to a larger-scale $z\sim 6\pm0.2$ overdensity reported by \citet{brinchCOSMOS20202023}, which may actually span multiple distinct substuctures, connected by filaments \citep{brinchDEIMOS2024}.

As is the case around HD850.1, recent \JWST/NIRSpec observations of the PRIMER-COSMOS field have greatly accelerated the confirmation of $z>5$ galaxies. 
In addition to the spectrum of \mambo-9-B presented in \S\ref{sec:nirspec}, we search all spectra from CAPERS for objects potentially associated with \mambo-9. 
We additionally searched DD\#6585 (P.I.~D.~Coulter), a cycle 3 DDT program targeting high-redshift supernovae candidates, and GO\#1810 (P.I.~S.~Belli), a cycle 1 program targeting massive galaxies at cosmic noon. 
Figure~\ref{fig:overdensity} shows the on-sky distribution of spectroscopically-confirmed galaxies between $z=5.5$ and $6.2$ (colored by redshift), including Keck/VLT observations as well as \JWST/NIRSpec. 
The inset panel shows the redshift histogram, revealing a clear excess of sources between $z=5.8$--$5.9$, confirmed via their strong \oiii\ and/or H$\alpha$ emission.

The structure appears to be quite large, spanning most of the PRIMER-COSMOS field, with $39$ objects within $\Delta z = 0.03$ of \mambo-9 (i.e., $z=5.82$--$5.88$). 
While NIRSpec/MSA observations inherently suffer from incompleteness, NIRCam/WFSS observations, primarily from COSMOS-3D (GO\#5893, P.I.~Kakiichi), will enable a more comprehensive mapping of the full protocluster and the filamentary structure as traced by H$\alpha$-emitters. 
If the structure indeed spans $\sim 40$\,cMpc, or is connected to another overdensity at $z\sim 6.0$ \citep{brinchDEIMOS2024}, spanning line-of-sight distances $\sim 100$\,cMpc, it would be among the largest protoclusters known. 
Understanding how massive galaxies grow may require characterizing the environments they live in. 

\section{Summary}\label{sec:summary}
We have presented new, high-resolution ALMA \cii\,158\,$\mu$m and dust continuum observations and \JWST/NIRCam+MIRI imaging of \mambo-9, a merging pair of dust star-forming galaxies at $z=5.85$.  
We detect \cii\ and dust continuum emission in \mambo-9-A and B. 
The dust continuum emission in both galaxies is concentrated in a central core, and \mambo-9-B exhibits a tail towards \mambo-9-A. 
The \cii\ emission is more diffuse, and both galaxies exhibit \cii$/L_{\rm FIR}$ ratios consistent with the ``\cii\ deficit'', in which \cii\ emission is weaker in region of higher FIR surface density. 
This trend holds in a spatially-resolved analysis, indicating that the \cii\ deficit arises from scales $\lesssim 400$\,pc. 

Both galaxies exhibit double-peaked \cii\ lines and clear velocity gradients, suggesting rotation. 
We fit the datacubes with \bbarolo\ and derive dynamical masses of $(10.6\pm 0.9)\times 10^{10}$\,M$_\odot$ for \mambo-9-A and $(2.0\pm 0.5)\times 10^{10}$\,M$_\odot$ for \mambo-9-B. 
The mass ratio of $\sim 1$:$5$ is consistent with a minor merger. 

New \JWST/NIRCam+MIRI imaging reveals complex morphology in the rest-frame optical/near-IR. 
\mambo-9-A and B are connected by a continuous bridge, itself quite red, possibly a tidal tail from a recent interaction. 
The central cores of both galaxies, where the dust continuum emission peaks, are faint in the rest-frame optical.
We conduct spatially-resolved panchromatic SED fitting in 2D bins across the extent of the \mambo-9 system; we find that while the bulk of the recent star-formation is concentrated in compact, highly obscured ($A_V\gtrsim 10$) clouds, the bulk of the optical light is emergent from moderately obscured ($A_V\sim1$--$5$) regions on the outskirts. 

Comparison of the total baryonic mass ($M_\star + M_{\rm gas} + M_{\rm dust}$) to the derived dynamical masses requires a low $\alpha_{\rm CO}$ ($\sim 1$) or gas-to-dust ratio ($\sim 80$), indicative of a highly metal-enriched ISM. 
\mambo-9 appears to reside in a large overdensity: we identify 39 galaxies spectroscopically confirmed within $\sim 25$ cMpc of \mambo-9, spanning most of the PRIMER-COSMOS field. 
Future spectroscopic efforts, including NIRCam WFSS observations from COSMOS-3D, will help reveal the connection between galaxy growth and environment.

\begin{acknowledgements}

H.B.A. acknowledges the support of the UT Austin Astronomy Department and the UT Austin College of Natural Sciences through Harrington Graduate Fellowship, as well as the National Science Foundation for support through the NSF Graduate Research Fellowship Program. 
CMC thanks the NSF for support through AST-2009577 and AST-2307006, as well as NASA through JWST-GO-01727.
JBC acknowledges funding from the JWST Arizona/Steward Postdoc in Early galaxies and Reionization (JASPER) Scholar contract at the University of Arizona.
AJT acknowledges support from the UT Austin College of Natural Sciences and from STScI/NASA through JWST-GO-6368
This paper makes use of the following ALMA data: ADS/JAO.ALMA\#2021.1.00505.S.
ALMA is a partnership of ESO (representing its member states), NSF (USA) and NINS (Japan), together with NRC (Canada), NSTC and ASIAA (Taiwan), and KASI (Republic of Korea), in cooperation with the Republic of Chile.
The Joint ALMA Observatory is operated by ESO, AUI/NRAO and NAOJ.
The National Radio Astronomy Observatory is a facility of the National Science Foundation operated under cooperative agreement by Associated Universities, Inc.

This work is based on observations made with the NASA/ESA/CSA \textit{James Webb Space Telescope}, obtained at the Space Telescope Science Institute, which is operated by the Association of Universities for Research in Astronomy, Incorporated, under NASA contract NAS5-03127. 
Support for program number GO-6368 was provided through a grant from the STScI under NASA contract NAS5-03127. 
The data were obtained from the Mikulski Archive for Space Telescopes (MAST) at the Space Telescope Science Institute.
These observations can be accessed via \dataset[doi: 10.17909/0q3p-sp24]{\doi{10.17909/0q3p-sp24}}.

Authors from UT Austin acknowledge that they work at an institution that sits on indigenous land. 
The Tonkawa lived in central Texas, and the Comanche and Apache moved through this area. 
We pay our respects to all the American Indian and Indigenous Peoples and communities who have been or have become a part of these lands and territories in Texas.

\end{acknowledgements}

\newpage
\bibliographystyle{aasjournalv7} 
\bibliography{MAMBO-9.bib}

\end{document}